\documentclass[letterpaper]{jpconf}

\usepackage{epsfig}
\begin{document}
\title{Studies of multiplicity in relativistic heavy-ion collisions}
\author{B.B.Back}
\address{Physics Division, Argonne National Laboratory, Argonne, IL 60439-4843, USA}

\ead{back@anl.gov}

\begin{abstract}
In this talk I'll review the present status of charged particle multiplicity measurements from heavy-ion collisions. The characteristic features of multiplicity distributions obtained in Au+Au collisions will be discussed in terms of collision centrality and energy and compared to those of p+p collisions. Multiplicity measurements of d+Au collisions at 200 GeV nucleon-nucleon center-of-mass energy will also be discussed. The results will be compared to various theoretical models and simple scaling properties of the data will be identified.   
\end{abstract}

\section{Introduction}

The multiplicity of charged particles emitted in a heavy-ion collision constitutes an important observable, which reflects the properties of the hot and dense system formed in the overlap region between the two incoming nuclei. Even without more detailed and differentiated measurements of the emitted particles one can obtain important information about the collision from measurements of the total multiplicity of charged particles, its distribution in pseudorapidity space (angular dependence) and its dependence on collision centrality and energy. Central questions concerning the redistribution of the incoming energy into particle production and kinetic energy can be addressed on the basis of relatively simple multiplicity measurements. It is therefore very appropriate that the topic of particle multiplicity in heavy-ion collisions is addressed in this workshop, and in this article I will attempt to give an overview of multiplicity measurements at RHIC energies and to compare these data with measurements at lower energies as well as simpler collisions between individual protons and leptons.  

\begin{figure}[bt]
\centerline{
\epsfig{file=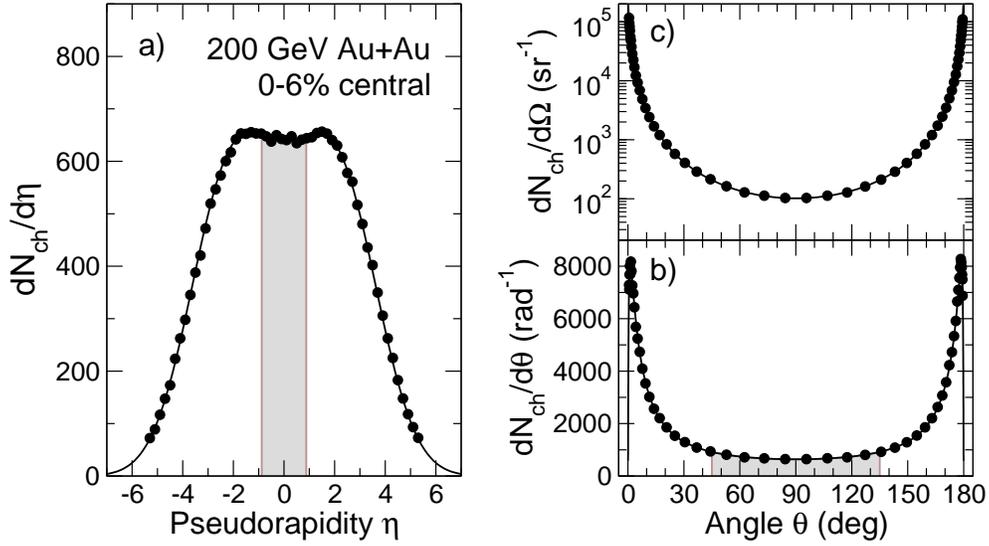,width=13cm}}
\caption{Illustration charged particle distribution for 0-6\% central 200 GeV Au+Au collisions~\cite{limfrag}. Panel a: The pseudorapidity density, $dN/d\eta$ is shown as a function of $\eta$. Panel b: The corresponding angular distribution $dN/d\theta$ is shown as a function of the angle $\theta$ relative to the beam axis. Panel c: same as for panel (b) but here $dN/d\Omega$ is shown. The shaded regions in panels a) and b) indicate the angular region where the transverse momentum $p_t$ exceeds the longitudinal momentum $p_{||}$.}
\label{dNdtheta_AuAu200}
\end{figure}

Some of the general properties of charged particle multiplicity distributions may be seen in Fig.~\ref{dNdtheta_AuAu200}. Here the distribution of charged particles, $dN_{ch}/d\eta$, where $\eta=-\ln\tan(\theta/2)$, is shown as a function of pseudorapidity for central $Au+Au$ collisions~\cite{limfrag} (solid points). The solid curve represents a triple-Gaussian fit to the data. One observes several distinct features of this distribution: an approximately flat region near mid-rapidity extends in this case out to $|\eta| \sim 2$ followed by a smooth fall-off region toward larger/smaller values of $\eta$. Since the particles emitted  within the angular region $45^\circ < \theta < 135^\circ$ corresponding to $-0.88 < \eta < 0.88$ (grey bands in panels a) and b)) are most likely to represent a thermalized region of phase space, there is special significance attached to the number of charged particles emitted in this region, {\it i.e.} the height of the mid-rapidity plateau or $dN_{ch}(|\eta|<1)/d\eta$. Although this region appears as a plateau in the $dN_{ch}/d\eta$-distribution shown in panel a) it is interesting to observe that the $dN_{ch}/d\theta$ and the $dN_{ch}/d\Omega$ distributions both exhibit a minimum at $\theta=90^\circ$ corresponding to $\eta=0$. In reality the charged particle distributions are very forward peaked.

\section{Mid-rapidity densities}

\begin{figure}[bt]
\centerline{
\epsfig{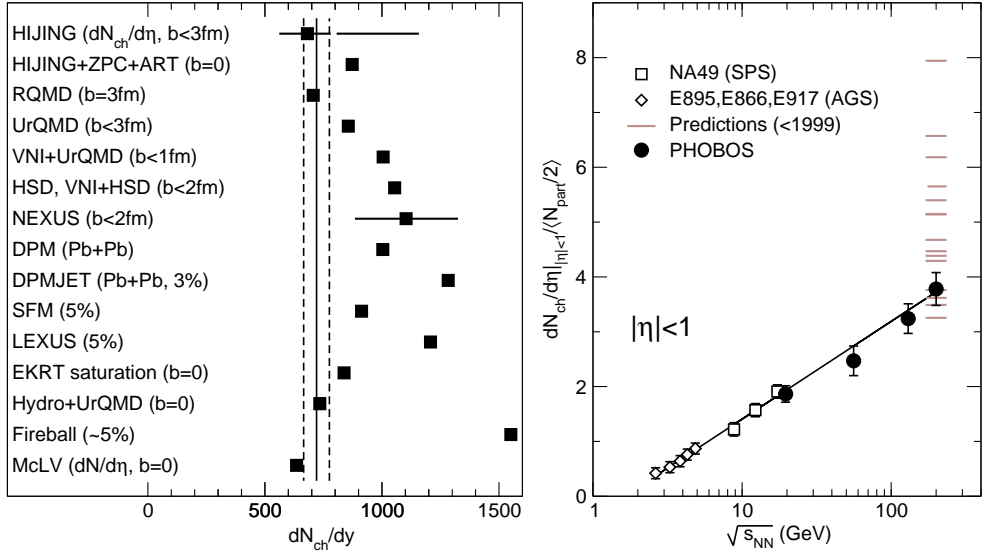}}
\caption{Left panel: A compilation of theoretical model predictions (solid squares) of the mid-rapidity density of charged particles $dN_{ch}/d\eta$ or $dN_{ch}/dy$ made prior to the start of the RHIC in July 2000 \cite{EskolaQM01}. The vertical line is then PHOBOS measurement~\cite{PH_20_200} at 200 GeV multiplied by a factor of 1.1 to correct for the $dN/d\eta$ to $dN/dy$ transformation. Right panel: The Energy dependence of the mid-rapidity $|\eta|<1$ pseudo-rapidity density. Data obtained at RHIC (solid circles)~\cite{PH_56_130,PH_20_200}, SPS (open squares)\cite{NA49}, and AGS (open diamonds)\cite{AGS} show a roughly logarithmic increase with collisions energy. $\sqrt{s_{\it NN}}$ (solid line) , but are also consistent with the trend of the Gluon Saturation Model (dashed curve) at $\sqrt{s_{\it NN}}>20$ GeV.} 
\label{dNdeta0_predict}
\end{figure}

Before the start of the RHIC program, several theoretical predictions had been made concerning the density of charged particles at mid-rapidity, $dN/dy|_{y=0}$ for $\sqrt{s_{\it NN}}$=200 GeV central $Au+Au$-collisions. These predictions are summarized in the left panel of Fig.~\ref{dNdeta0_predict} which has been adapted from a compilation by Eskola \cite{EskolaQM01}. The vertical band represents the PHOBOS measurement of $dN/d\eta|_{|\eta|<1}$ multiplied by a factor of 1.1 \cite{EskolaQM01} to correct for the transformation to rapidity, $y$. It is evident that most of the predictions overestimated the density by up to a factor of two although a few predictions agree with the measurement.

Measurements of $dN_{ch}/d\eta$ may also be used to obtain an estimate of the energy density achieved in the collision region. A simple estimate may be based on the naive assumption that the total available energy in 200 GeV $Au+Au$ collisions of $\sim$39 TeV could be converted into a completely stopped source of high energy density that emits particles isotropically in the final state. This leads to an energy density of $\sim$ 130 GeV/fm$^3$, based on a formation time of $\tau_0\sim$ 1 fm/c. Since this estimate exceeds the critical energy density for the transition into a partonic state (QGP) by two orders of magnitude, one may consider it inevitable that this new state of matter is formed in such collisions. Studies at lower collision energies performed at the AGS and SPS have shown, however, that this complete stopping picture is invalid. 
Also the present measurements are at variance with this picture. The  $dN_{ch}/d\eta$ distributions observed are substantially wider than the  $dN_{ch}/d\eta \propto 1/\cosh^2\eta$ shape corresponding to the fully stopped / isotropic source scenario. 
Alternatively, the energy density may be estimated more reliably from the method proposed by Bjorken~\cite{Bjorken}, which is based on the total energy of particles emitted at mid-rapidity. Espressed in terms of the $dN_{ch}/d\eta$ distribution we find
\begin{equation}
\epsilon_0=\frac{\langle m_t \rangle}{\pi R^2 \tau_0}\frac{dN_{ch}}{d\eta}f_{\it neut}f_{\it Jac}, 
\end{equation}
where $\langle m_t \rangle$ is the mean transverse mass, $\pi R^2$ is the transverse area of the fireball, $\tau_0$ is formation time and  the factors $f_{\it neut}$ and $f_{\it Jac}$ correct for the unobserved neutral particles and the Jacobian in the pseudorapidity to rapidity transformation. Inserting values of $\langle m_t \rangle$=0.57 GeV/c$^2$, $f_{neut}$=1.6, $f_{Jac}$=1.1, all of which are obtained from particle identified spectra measured by BRAHMS~\cite{BRAHMS_PiK}, using $dN_{ch}/d\eta$=700 for the 3\% most central $Au+Au$ collisions and a formation time of $\tau_0$=1 fm/c one finds an energy density of $\epsilon_0 \approx$5 GeV/fm$^3$ , which is still substantially above the predicted threshold for the phase transition of $\epsilon_0\sim$0.7-1.0 GeV/fm$^3$ obtained from lattice QCD calculations~\cite{Karsch} . Much of the available energy is thus carried off by particles emitted in forward-backward angles associated with lower energy density, which gives rise to pseudorapidity distributions that are substantially broader than expected for an isotropic source. The pseudorapidity distribution is therefore intimately connected to the energy density of the emitting source and provides an important test-bed for validating or discrediting theoretical models attempting to describe the conditions in the early phases of the collision.

The dependence on collision energy is shown in the right panel of Fig.~\ref{dNdeta0_predict}. The mid-rapidity charged particle density $dN_{ch}(\eta=0)/d\eta/\langle N_{part}/2\rangle$ is divided by the number of participant pairs $N_{part}/2$ in order correct for the small difference in the system size between $Au+Au$ and $Pb+Pb$ collisions. The data include fixed target measurements of $Au+Au$ collisions at the AGS~\cite{AGS} at Brookhaven National Laboratory and $Pb+Pb$ collisions at the SPS at CERN~\cite{NA49} as well as colliding $Au$-beam collisions at four RHIC energies. First it is interesting to note that the measurements at $\sqrt{s_{\it NN}}=200$ GeV (the highest energy) falls in the lower part of the pre-2000 predictions (grey bars). This discrepancy between many of the predictions led to a strong revision of several model assumptions and discarded models which led to a severe over prediction of this simple observable.

\begin{figure}[bt]
\centerline{
\epsfig{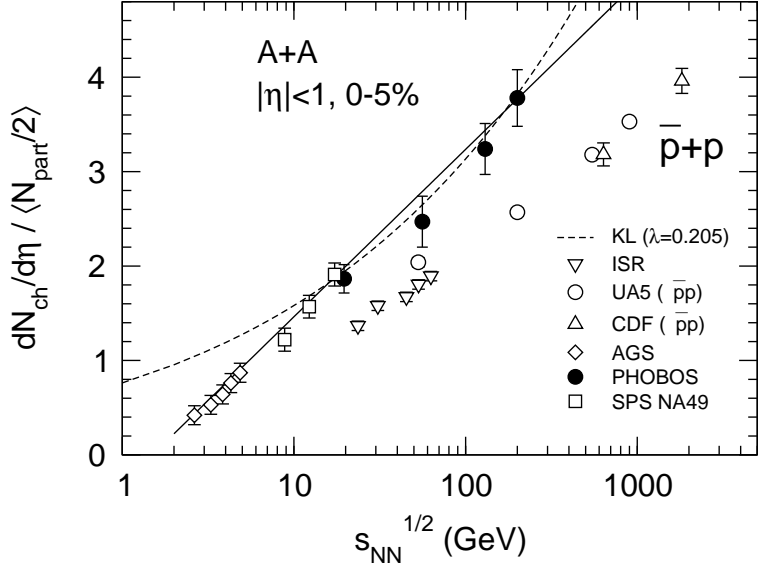}}
\caption{ The energy dependence of the mid-rapidity $|\eta|<1$ pseudo-rapidity density for central heavy-ion collisions. Data obtained at RHIC (solid circles)~\cite{PH_56_130,PH_20_200}, SPS (open squares)\cite{NA49}, and AGS (open diamonds)\cite{AGS} show a roughly logarithmic increase with collisions energy, $\sqrt{s_{\it NN}}$ (solid line) , but are also consistent with the trend of the Gluon Saturation Model (dashed curve) at $\sqrt{s_{\it NN}}>20$ GeV. Corresponding data from $pp$~\cite{Thome} and $\overline{p}p$~\cite{Alner,Abe} collisions are shown as open symbols}.
\label{dNdeta_eta0}
\end{figure} 

In addition, one observes that the energy dependence rather closely approximates a logarithmic increase denoted by the solid curve such that 
\begin{equation}
\frac{2}{\langle N_{\it part}\rangle}\frac{dN_{ch}|_{|\eta|<1}}{d\eta} = 
0.77\ln\sqrt{s_{\it NN}}-0.31.
\label{ln(s)}
\end{equation}
None of the present models reproduce this exact energy dependence. However, the gluon saturation model of Kharzeev, Levin, and Nardi~\cite{KN,KL} agrees rather well with the data above $\sqrt{s_{\it NN}}>20$ GeV as indicated by the dashed curve in Fig.~\ref{dNdeta_eta0}. This model leads to the expression
\begin{equation}
\frac{1}{N_{\it part}/2}\frac{dN_{\it ch}}{d\eta}=2c\left(\frac{s}{s_0}\right)^{\lambda/2} \ln \left(\frac{Q_s^2}{\Lambda^2_{\it QCD}}\right)/\sqrt{1+\frac{m^2}{p_t^2}}.
\label{eq10}
\end{equation}

Here, $Q_s$ is the gluon saturation momentum, $\Lambda_{\it QCD}$ is the QCD scale parameter, $c$ is a normalization constant that is fixed at a single energy of $\sqrt{s_0}$=130 GeV, and $\lambda$ provides a scaling with energy, which for HERA data assumes a value of $\lambda \simeq 0.25-0.3$ \cite{KL}
The dashed curve shown in Fig.~\ref{dNdeta_eta0} corresponds to the following values; $c=1.02$, $\lambda=0.205$ (obtained from the fit to the data), $Q_s^2=2.02(s/s_0)^{\lambda/2}$\cite{KN},   $m=0.75\sqrt{Q_s/c \times 1 {\rm GeV/c^2}}$, and $p_t=Q_s$. Although some of these values differ slightly from those listed in Ref.~\cite{KN}, they are all considered to be in a reasonable range. 

In addition, one observes from Fig.~\ref{dNdeta_eta0} that the particle production in heavy-ion collisions (solid symbols) is substantially higher that what is seen in $pp$~\cite{Thome} or $\overline pp$~\cite{Alner,Abe} collisions (open symbols) when normalized to colliding nucleon-nucleon pair, $N_{part}/2$. One possible interpretation of this difference is that a large fraction ($\sim$50\%) of the available energy is carried off by leading hadrons in $pp$ and $\overline pp$ collisions, whereas this effect is absent in heavy-ion collisions because these hadrons suffer subsequent collisions leading to particle production in the heavy-ion collision environment~\cite{universality}    

\section{Centrality dependence}

\begin{figure}[bt]
\centerline{\epsfig{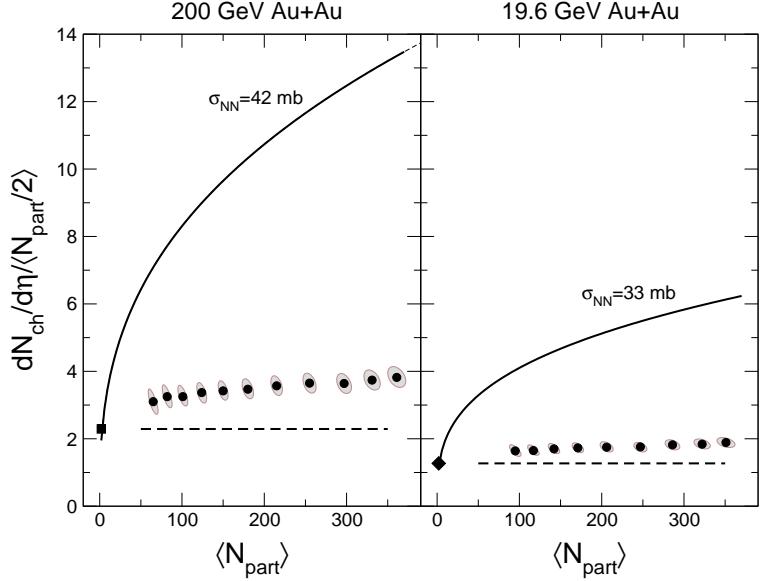}
}
\caption{The mid-rapidity density, $dN_{\it ch}/d\eta$ scaled to the number of participant pairs, $\langle N_{part}/2 \rangle$ is shown as a function of centrality, $\langle N_{part} \rangle$ for 200 GeV (left panel) and 19.6 GeV (right panel)~\cite{PH_20_200}. The solid squares (diamonds) represent the $\overline pp$ ($pp$) values at 200 (19.6) GeV, respectively \cite{Alner,Thome}. The solid curves show the scaling with the number of collisions obtained by the Glauber model using nucleon-nucleon cross sections of 42 mb and 33 mb for 200 and 19.6 GeV, respectively}  
\label{Npart_Ncoll}
\end{figure}
In heavy-ion collisions it is possible to obtain additional information on the particle production by varying the impact parameter between the two colliding ions, which also varies the size of the collision volume. In all four RHIC experiments, the centrality of a collision is measured on the basis of the total charged particle multiplicity or energy with in a certain pseudo-rapidity region. The methods for obtaining the centrality in the PHOBOS experiment is described in detail in a separate contribution to these proceedings~\cite{BariHollis} and will not be  discussed further here.

As examples of the centrality dependence, Fig.~\ref{Npart_Ncoll} shows the mid-rapidity density, $dN/d\eta|_{|\eta|<1}/\langle N_{part}/2\rangle$ (solid points)  as a function the number of participants, $N_{part}$ for $Au+Au$ collisions at 19.6 and 200 GeV~\cite{PH_20_200}. One observes a steady increase in the particle production with centrality in both cases, whereas scaling with the number of participants would lead to a flat dependence. This is shown by the dashed lines which correspond to the level seen in nucleon-nucleon collisions \cite{Alner} (200 GeV $\overline pp$), \cite{Thome} (19.6 GeV extrapolated $pp$). On the other hand, if the particle production would scale with the number of inelastic collisions between nucleons (obtained from Glauber model simulations), one would expect a very strong centrality dependence as shown by the solid curves in Fig.~\ref{Npart_Ncoll}. Clearly, this strong dependence is not seen in the data; they are much closer to the $N_{part}$ scaling limit.

\begin{figure}[bt]
\centerline{\epsfig{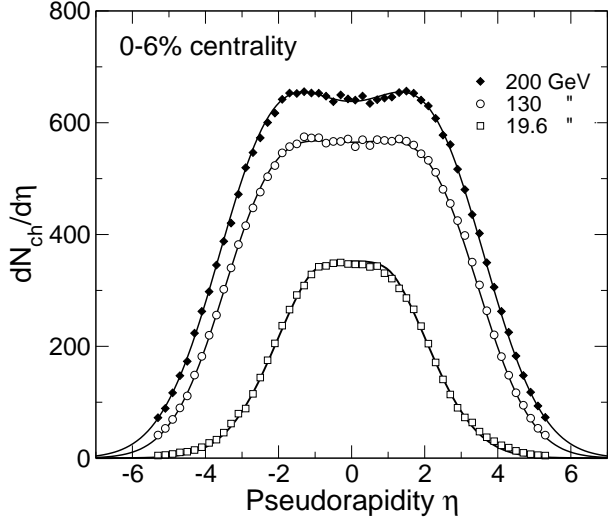}}
\caption{The shapes of the pseudorapidity density, $dN_{\it ch}/d\eta$, is shown as a function of $\eta$ for 0-6\% central Au+Au collisions at 19.6 (open squares), 130 (open circles), and 19.6 GeV (solid diamonds)~\cite{limfrag}. The solid curves represent best fits to the data using triple Gaussians.}  
\label{central_20_130_200}
\end{figure}

\section{$dN/d\eta$ Shapes}

The energy evolution of the shape of $dN/d\eta$ distributions in central $Au+Au$-collisions is illustrated in Fig.~\ref{central_20_130_200}. Several observations can be made: 1) The mid-rapidity plateau increases in both width and height with collision energy and 2) the extend of the fall-off regions outside of this plateau also increases with energy, but the slope of the fall-off is essentially independent of energy, a consequence of limiting fragmentation scaling~\cite{limfrag}. At first glance it may appear strange that an order of magnitude increase in the collision energy results in only about a factor of two increase in the mid-rapidity value of $dN_{ch}/d\eta$. Where does the energy go? Here it is worth keeping in mind that the energy of a particle is $E=\sqrt{m_0^2+p_t^2\cosh^2\eta}$, where $m_0$ and $p_t$ is its rest mass and transverse momentum, respectively. The energy per particle therefore increases sharply with $\eta$ provided that $\langle p_t \rangle$ does not fall off precipitously away from $\eta$=0 . Thus, the energy of a pion with $p_t$=0.5 GeV emitted at the half maximum point of $\eta \sim$ 2 at 19.6 GeV is $E$=1.9 GeV whereas it requires an energy of $E$=11.2 GeV to emit the same particle at $\eta \sim$3.8, the half maximum at 200 GeV. The increased width of the $dN_{ch}/d\eta$ distribution therefore consumes the largest fraction of the additional energy.

\begin{figure}
\centerline{\epsfig{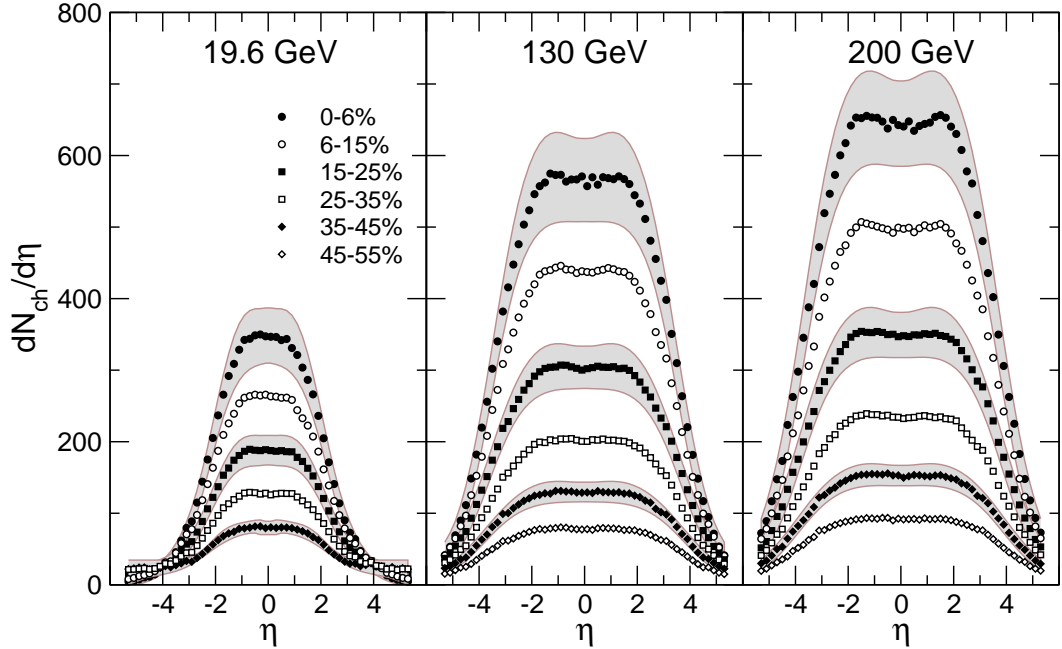}}
\caption{$dN_{\it ch}/d\eta$ vs $\eta$ (solid and open points) for five centrality bins representing 45\% of the total cross section for $\sqrt{s_{\it NN}}$=19.6 GeV $Au+Au$ collisions and for six centrality bins for $\sqrt{s_{\it NN}}$=130 and 200 GeV correcponding to 55\% of the cross section~\cite{limfrag}. The shaded bands represent estimated systematic errors.}
\label{Fig_211}
\end{figure}

The evolution with both energy and centrality is shown in Fig.~\ref{Fig_211} for $Au+Au$ collisions measured by PHOBOS~\cite{limfrag}. Similar data at 130 and 200 GeV have been obtained by the BRAHMS collaboration~\cite{Brahms130,Brahms200}

\begin{figure}
\centerline{\epsfig{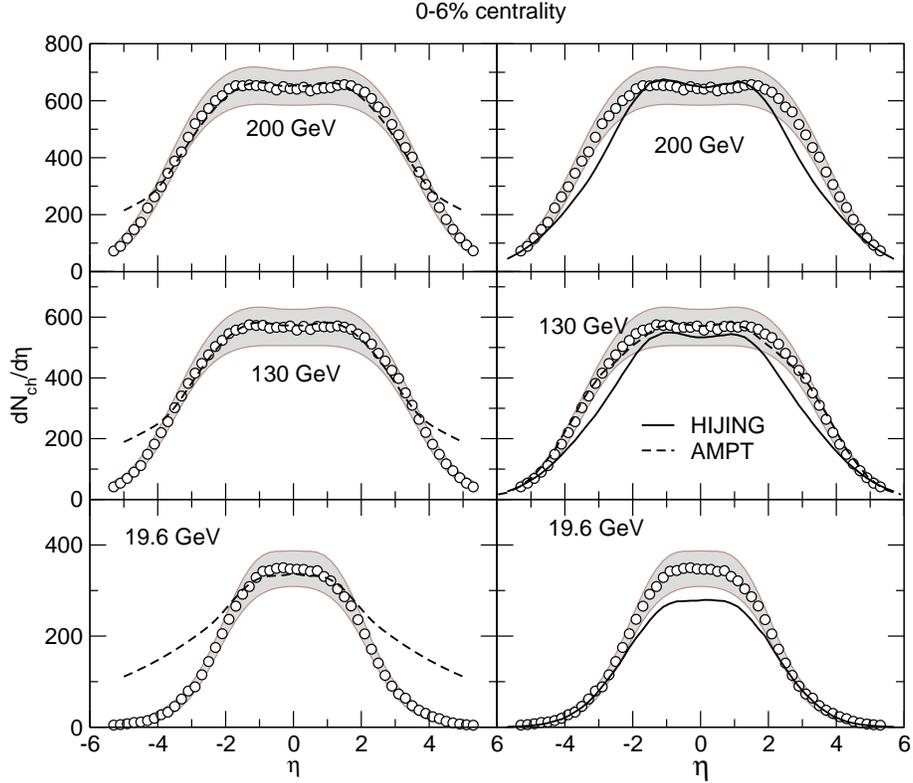}}
\caption{Pseudorapidity distributions of charged particles emitted in central (0-3\% bin) Au+Au collisions at $\sqrt{s_{\it NN}}$= 19.6, 130, and 200 GeV~\cite{limfrag} (symbols) are compared with saturation model calculations, Eqs. 7-9 (dashed curves). The solid curves represent multi-Gaussian fits to the data, see text.Pseudorapidity distributions of charged particles emitted in central (0-3\% bin) Au+Au collisions at $\sqrt{s_{\it NN}}$= 19.6, 130, and 200 GeV (symbols) are compared with saturation model calculations, Eqs. 7-9 (dashed curves). The solid curves represent multi-Gaussian fits to the data, see text. are compared to calculations with the HIJING~\cite{HIJING} (solid curves) and AMPT~\cite{AMPT} (dashed curve) models.}
\label{dNdeta_models}
\end{figure}

In Fig.~\ref{dNdeta_models} central $dN_{ch}/d\eta$ distributions at 19.6, 130 and 200 GeV are compared with model calculations. The dashed curves shown in the left panels are obtained from the gluon saturation model of Kharzeev {\it et al.}~\cite{KN,KL} using the analytical expression
\begin{eqnarray}
\label{Saturation}
\nonumber \frac{dN}{dy}=cN_{\it part}\left(\frac{s}{s_0}\right)^{\lambda/2} e^{-\lambda|y|}\left[\ln\left(\frac{Q_s^2}{\Lambda^2_{\it QCD}}\right)-\lambda|y|\right]\\
\times\left[1+\lambda|y|\left(1-\frac{Q_s}{\sqrt{s}}e^{(1+\lambda/2)|y|}\right)^4\right].
\end{eqnarray}
and the parameter values listed in Sect. 2. Note that a rigorous transformation from $y$ to $\eta$ requires an integration over the actual $p_t$ distribution for each particle species, {\it i.e.} pions, kaons, and nucleons, which is not carried out here. We find that this analytical expression gives a reasonable account of the $dN_{ch}/d\eta$ distributions over most of the $\eta$ range, but that it deviates at the largest pseudorapidities, where the approximations on which it is based, are not fulfilled. In fact, a more accurate calculation by Kharzeev {\it et al.}\cite{KLN} indicate that also the 19.6 GeV data can be reproduced to a satisfactory degree.   

The 0-6\% most central data are also compared to the HIJING~\cite{HIJING} AMPT~\cite{AMPT} models as shown in the right hand panels of Fig.~\ref{dNdeta_models}. It is evident that the HIJING model (solid curves) tend to underpredict the width of the distributions, whereas the AMPT model calculation at 130 GeV appears to alleviate this discrepancy. Since the AMPT model uses the HIJING event generator but includes the final state interactions, it appears that this effect is important in the calculations.   

\subsection{Boost invariance?}

\begin{figure}
\centerline{\epsfig{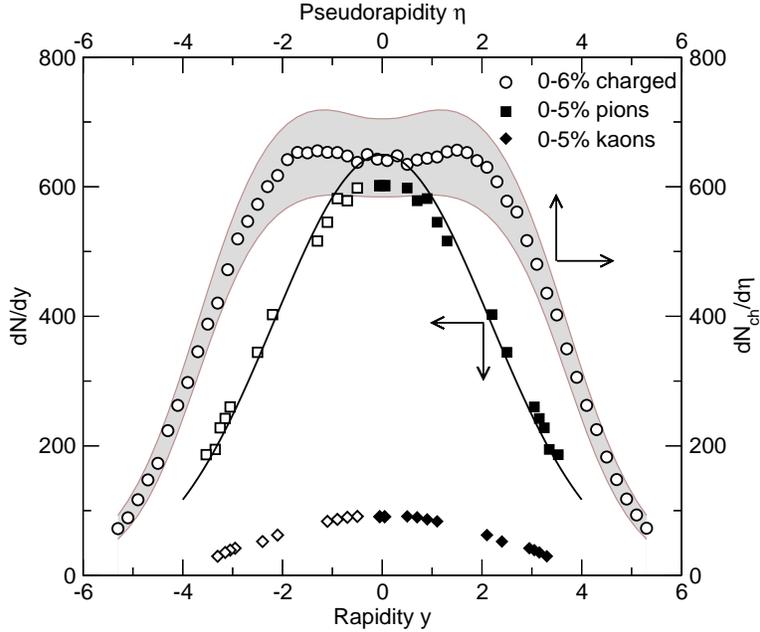}}
\caption{Visual comparison of the $dN/dy$ distributions for pions (squares) and kaons (diamonds) measured by BRAHMS~\cite{BRAHMS_PiK} with the $dN_{ch}/d\eta$ distribution for all charged particles measured by PHOBOS~\cite{limfrag}  0-5\% and 0-6\% central Au+Au collisions at 200 GeV, respectively. Note the different axes used for the two different types of data. The solid curve represents a Gaussian with $\sigma$=2.16 (see text).}  
\label{BRAHMS_pionsKaons}
\end{figure}

Initially it was thought that a boost invariant region around mid-rapidity would develop at sufficiently high collision energy~\cite{Bjorken}. At first sight one might interpret the flat-top shape of $dN_{ch}/d\eta$ distributions shown in Fig.~\ref{central_20_130_200} as evidence for boost invariance, but this plateau is only a consequence of the transformation from rapidity to pseudurapidity space in which the Jacobian gives rise to a reduction of $dN_{ch}/d\eta$ near mid rapidity. This effect is clearly illustrated in Fig.~\ref{BRAHMS_pionsKaons}. Here the $dN_{ch}/d\eta$ distribution for 0-6\% central 200 GeV $Au+Au$ collisions measured by PHOBOS~\cite{limfrag} (open circles) is compared with the $dN/dy$ distributions of pions and kaons measured by BRAHMS for 0-5\% centrality for the same collision system. Whereas the $dN_{ch}/d\eta$ distribution exhibit a flat-top shape, a nearly Gaussian shape is observed for the $dN/dy$ distribution, the width of which is in rather good agreement with the predictions of the hydrodynamical expansion model proposed by Landau~\cite{Landau}. In this model it is assumed that full stopping is achieved between the two colliding ions followed by isentropic expansion of a thermally equilibrated system. This leads to an approximately Gaussian shape of the $dN/dy$ distribution \cite{Milekhin}. In a simplified version of this model \cite{Carruthers} the standard deviation of the distribution is given by $\sigma^2=\ln(\sqrt{s_{\it NN}}/2m_p)$, where $m_p$ is the proton mass. In the case of 200~GeV $Au+Au$ collisions, $\sigma=2.16$, the value used for calculating the solid curve in Fig.~\ref{BRAHMS_pionsKaons}. The measured $dN/dy$ distribution of pions is seen to deviate from this curve only slightly in the mid-rapidity region which may indicate a slight tendency towards boost invariance. Overall, however, the data show that the simplified picture of a wide boost invariant region at mid-rapidity does not appear to be valid.  

\subsection{d+Au collisions}

\begin{figure}
\hspace{3cm}
\epsfig{file=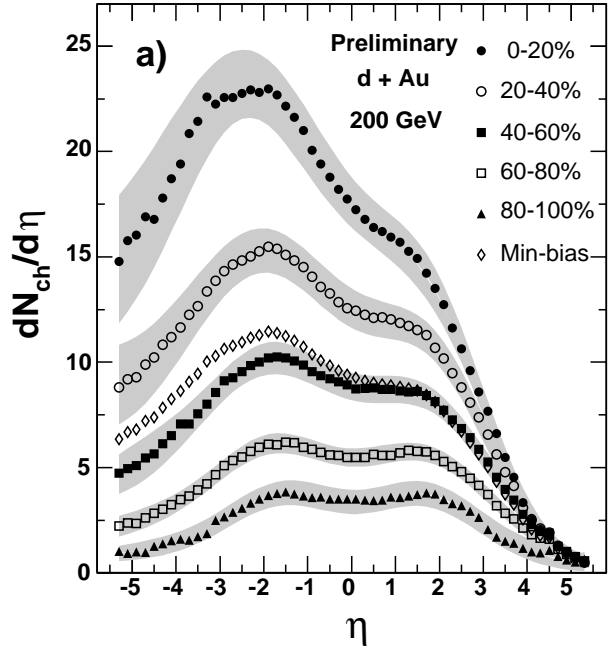,width=8cm}
\caption{The $dN_{ch}/d\eta$ distributions for $d+Au$ collisions at $\sqrt{s_{\it NN}}=$200 GeV as shown for different centrality bins~\cite{dAu_centrality}. The grey bands represent statistical and systematic errors. The minimum bias distribution is shown as open diamonds.}
\label{Fig15_dAu_White_paper}
\end{figure}

In an attempt to understand the effects of secondary collisions in the $Au+Au$ system, collisions between deuterons and $Au$ ions have been studied at RHIC. It was necessary to use deuterons instead of protons in order to better match the mass to charge ratio of the fully stripped $Au$-ions as both beams are bent in the same magnetic field near the interaction regions of the collider. In Fig.~\ref{Fig15_dAu_White_paper} the $dN_{ch}/d\eta$ distributions are shown for both minimum-bias $d+Au$ collisions (open diamonds) and different centrality bins given as percentage of the total inelastic cross section~\cite{dAu_centrality}. The method of determining the collision centrality is discussed in detail in Ref.~\cite{BariHollis}. Here the positive pseudorapidity corresponds to the deuteron beam direction whereas the $Au$ ions go in the negative $\eta$ direction. It is evident that the most abundant particle production is found at negative $\eta$ values for both the minimum-bias distribution as well as for the distributions of all but the most peripheral collisions. It is worth noting, however, that the reduction (or dip) near $\eta$=0 is most likely just an effect of the transformation to pseudorapidity space; it is expected that this reduction is absent in $dN/dy$ distributions for individual particle species. 

The asymmetric character of the $dN_{ch}/d\eta$ distributions is a natural consequence of longitudinal momentum conservation in the collision. Thus for a central collision, one may expect that the two deuteron participants (one neutron and one proton) interacts with more than ten nucleons in the $Au$ ion. The charged particles created in such a collision must therefore reflect the net momentum of the participants, which corresponds to $p_{||}=(N_{part}^{Au}-N_{part}^d)\times200$GeV/c.
This large excess of momentum in the $Au$ direction therefore naturally leads to a more abundant emission of particles in the negative pseudorapidity region. The longitudinal momentum asymmetry is, however, strongly reduced for peripheral collisions where in the 80-100\% bin we may expect that $N_{part}^{Au} \sim N_{part}^d$ and, indeed, we observe an almost symmetric $dN_{ch}/d\eta$ distribution in this case (solid triangles in Fig.~\ref{Fig15_dAu_White_paper}).

\subsection{Shapes of "elementary" collisions}

\begin{figure}
\centerline{\epsfig{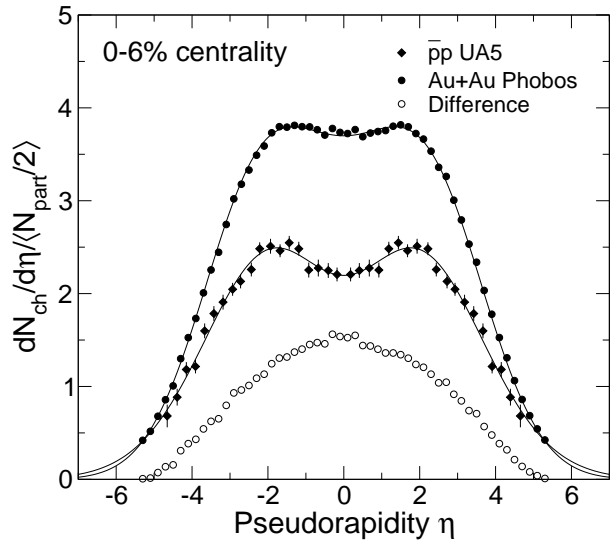}}
\caption{Comparison of the $dN_{ch}/d\eta/\langle_{part}/2\rangle$ distribution for 0-6\% central $Au+Au$ collisions at$\sqrt{s_{\it NN}}$=200 GeV~\cite{limfrag} (solid circles) and 200 GeV $\overline{p}p$~\cite{Alner} (solid diamonds) collisions. The difference between the two distributions is shown as open circles.}
\label{200AuAu_pp_shape}
\end{figure}

In Fig.~\ref{200AuAu_pp_shape} we compare the $dN_{ch}/d\eta$ shapes of central (0-6\%) $Au+Au$ (solid circles)~\cite{limfrag} and $\overline pp$\cite{Alner} (solid diamonds) collisions at $\sqrt{s_{\it NN}}$=200 GeV. The $Au+Au$ data have been normalized by the number of participant pairs $\langle N_{part}/2\rangle$. We observe that the additional particle production seen in $Au+Au$ collisions (open circles) is peaked at $\eta$=0. It is thus reasonable to associate this additional production with secondary collisions in the $Au+Au$ system because collisions tend to build up the transverse momentum component as the system progresses toward a thermal equilibrium. An overall increase of about 40\% is seen in the $Au+Au$ system. 

Although Fig.\ref{200AuAu_pp_shape} appears to indicate that the additional particle production in concentrated in the mid-rapidity region, the systematic errors associated with both measurements do also allow for an interpretation in which the additional particle production occurs over the while rapidity range with a constant enhancement factor of approximately 1.3~\cite{universality}.

It has also been noted that leptonic collisions give rise to $dN_{ch}/dy_T$ distributions (where $y_T$ is the particle rapidity along the thrust axis of the outgoing particle jets) that are similar to those of $pp/\overline{p}p$ and heavy-ion collisions~\cite{universality}. In this comparison it should, however, be kept in mind that some distortions to the $dN_{ch}/dy_T$ distribution will occur if they are converted into pseudorapidity space as illustrated above.

\subsection{Limiting fragmentation scaling}

\begin{figure}
\centerline{\epsfig{file=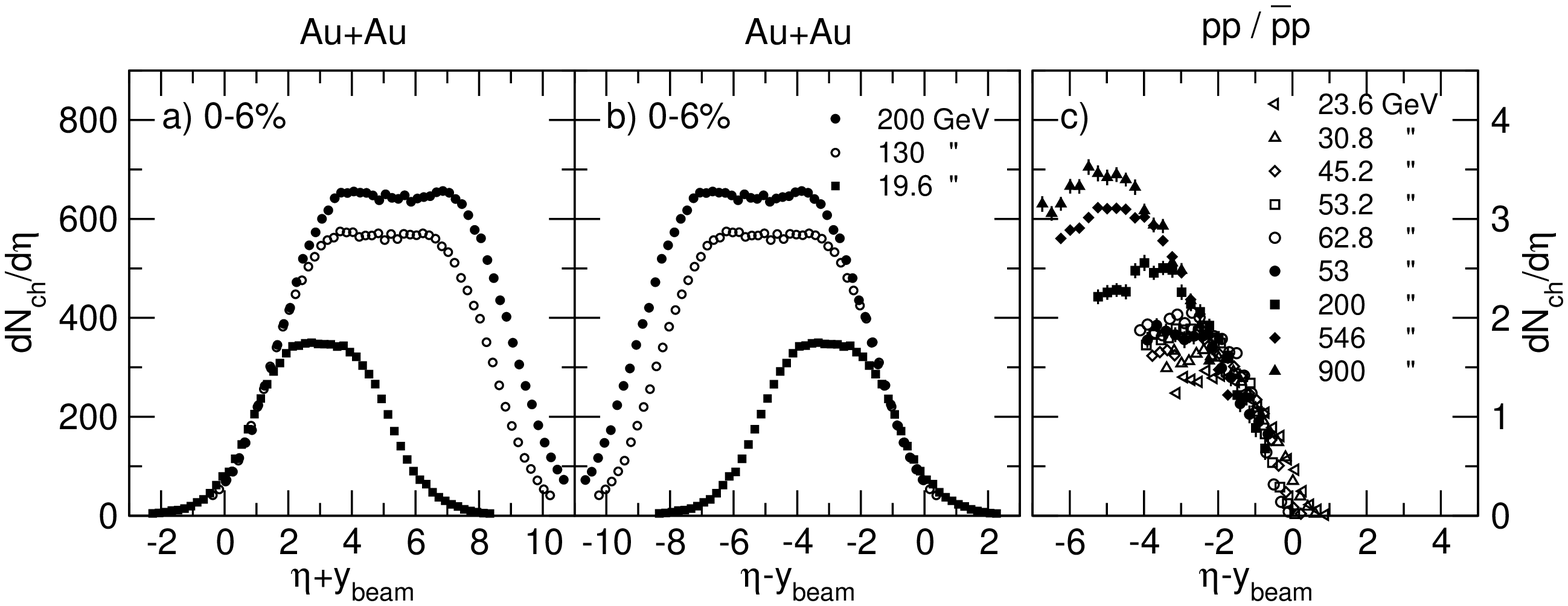,width=\textwidth}}
\caption{Limiting fragmentation scaling for 0-6\% central $Au+Au$~\cite{limfrag} and $pp/\overline{p}p$~\cite{Alner,Thome} collisions.}
\label{limfrag_AuAu_pp}
\end{figure}

In a previous section we discussed the observation that the width of the $dN_{ch}/d\eta$ distribution increases with collision energy. This effect is a consequence of limiting fragmentation scaling, according to which $dN/dy$ distributions are identical in the fragmentation region close to the target/beam rapidity region, {\it i.e.}, $dN/d(y-y_{beam})$ is energy independent \cite{Benecke}. Because for large $\eta$ values $y\simeq\eta+\ln(p_t/m_t)$, where the second term gives rise to only a small and essentially constant shift, one may also expect this scaling behaviour to apply to distributions measured in pseudorapidity space. This behavior has been observed first in $\overline{p}p$ collisions \cite{Alner}. In Fig.~\ref{limfrag_AuAu_pp}c, we illustrate that this scaling extends to substantially lower energies by including $pp$ data from Thome {\it et al.}\cite{Thome}. 

With the high-quality $dN_{ch}/d\eta$ data obtained by PHOBOS for $Au+Au$ collisions
it has been shown~\cite{limfrag} that limiting fragmentation scaling also holds very accurately for heavy-ion collisions as illustrated for central collisions in Fig.~\ref{limfrag_AuAu_pp}a,b. The charged particle production thus increases toward mid-rapidity with the same rate of about $\alpha=d^2N_{ch}/d\eta^2=195$ independent of collision energy.

\section{Total multiplicity}

As observed on Figs.~\ref{central_20_130_200} and \ref{Fig_211} the acceptance of the PHOBOS setup covers essentially the full distribution of charged particles emitted from $Au+Au$ collisions, even at the highest energy of $\sqrt{s_{\it NN}}$=200 GeV. The unobserved fraction may be estimated in several ways~\cite{universality}, namely by integrating fitted Woods-Saxon curves or by making use of the limiting fragmentation scaling to estimate the tails of the distributions. Several such methods give almost identical results. Thus it is found that the extrapolated region accounts for less than 1\% of the total for the most central bin increasing to $<5\%$ for peripheral collisions at 130 and 200 GeV. For 19.6 GeV it is believed that 100\% of the charged particles originating from the collision region fall within the detector acceptance. In this case the total number of charged particles, $N_{ch}$, was estimated by integrating over the PHOBOS acceptance, {\it i.e.} $|\eta|<5.4$. It is however, believed that the tails seen at large pseudorapidities, especially in peripheral collisions, are associated with emission from the spectators. This method may therefore lead to a slight over estimate of $N_{ch}$ for the most peripheral collisions.

\begin{figure}
\centerline{\epsfig{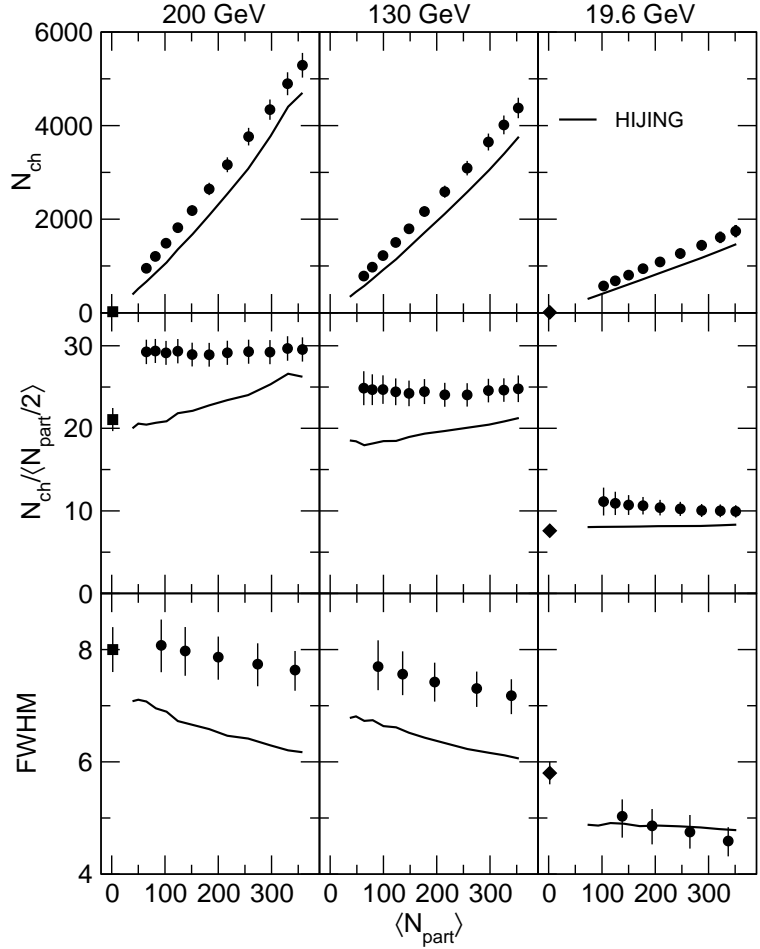}}
\caption{Upper panels: The total charged particle multiplicity~\cite{universality} (solid points) is shown as a function of centrality, $\langle N_{\it part}\rangle$ for three energies. Middle panels: Same as above but normalized to $\langle N_{\it part}\rangle/2$. Lower panels: Centrality dependence of the full width at half maximum (FWHM) of the $dN_{ch}/d\eta$ distributions~\cite{limfrag} (solid points). Solid curves represent HIJING calculations~\cite{HIJING}, whereas solid squares correspond to $\overline{p}p$~\cite{Alner} collisions. Solid diamonds represent an extrapolation to 19.6 GeV of $pp$ data~\cite{Thome}.}
\label{Phmult_fig16}
\end{figure}

The results are summarized in Fig.~\ref{Phmult_fig16} where the total number of charged particles, $N_{ch}$, is shown as a function of centrality expressed in terms of  $\langle N_{part} \rangle$ in the three upper panels (solid points)~\cite{universality}. The data are compared to HIJING calculations (solid curve) and to $\overline{p}p$ (solid square) \cite{Alner} and an extrapolation of $pp$ (solid diamond) \cite{Thome} collisions. By normalyzing to the average number of participants pairs $\langle N_{part} /2 \rangle$ an interesting trend becomes evident; unlike the mid-rapidity density per participant pair $dN/d\eta|_{|\eta|<1}/\langle N_{part}/2\rangle$ shown in Fig.~\ref{Npart_Ncoll}, which increases substantially with centrality, we find that the total normalized multiplicity remains constant with centrality within experimental errors. This requires that the width of the distributions become narrower with centrality in order to keep the total area constant. This trend is indeed observed in the bottom three panels of Fig.~\ref{Phmult_fig16} which show that the full width at half maximum, FWHM, of the distributions~\cite{limfrag} decrease with cenrality. It is interesting to note that the constant value of $N_{ch}/\langle N_{part}/2\rangle$ occurs at a substantially higher level than the $\overline{p}p$ data point, and that the HIJING model does not predict this constancy with centrality.

\begin{figure}
\centerline{\epsfig{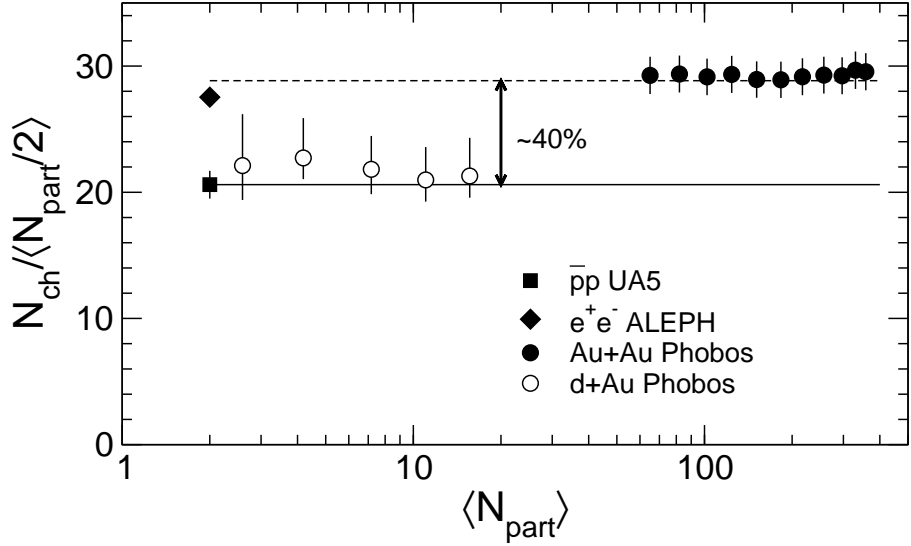}}
\caption{$N_{ch}/\langle N_{part}/2\rangle$ for $Au+Au$~\cite{universality} (solid points), $d+Au$~\cite{dAu_centrality} (open circles) are shown as a function of $\langle N_{part}\rangle$ and compared to valued for $\overline{p}p$~\cite{Alner} (solid square) and $e^+e^-$~\cite{e+e-} (solid diamonds) collisions. The horizontal dashed line represents the average value for $Au+Au$ collisions, which is $\sim$40\% higher than the $\overline{p}p$ level (solid line).}
\label{Nch_Npart_all}
\end{figure}

It is also of interest to determine whether the total charged particle production in the $d+Au$ collisions  shows the enhancement over simple nucleon-nucleon collisions seen for heavy ion collisions. The total number of charged particles is in this case estimated by an extrapolation outside the measured region (shown in Fig.~\ref{Fig15_dAu_White_paper}) based on the limiting fragmentation scaling using lower energy measurements of proton induced collisions~\cite{dAu_centrality}. The total normalized multiplicity, $N_{ch}/\langle N{\it part}/2\rangle$ for $d+Au$ is in Fig.~\ref{Nch_Npart_all} shown as a function $\langle N_{part} \rangle$ (open points) and compared to the data for $Au+Au$ (solid points), $\overline{p}p$ \cite{Alner} (solid square) and $e^+e^-$ \cite{e+e-} (solid diamond) collisions. We observe that the also the $d+Au$ multiplicity is substantially lower than for the heavy-ion system as was seen in the comparison to $\overline{p}p$ collisions. This effect does not appear to be simply a consequence of the number of participants in the collisions because there in no indication of an increased multiplicity even for the most central $d+Au$ collisions. We also note that the multiplicity for $e^+e^-$ collisions is consistent with the heavy-ion data. As mentioned earlier, one possible explanation for this disparity between nucleon-induced and heavy-ion or lepton induced collisions is the fact that leading particles in nucleon-induced collisions carry away about half of the energy, which is then not available for particle production~\cite{universality}. It is surprising, however, that this mechanism is still effective in $d+Au$ collisions, which involve a substantial number of subsequent collisions.

\begin{figure}
\centerline{\epsfig{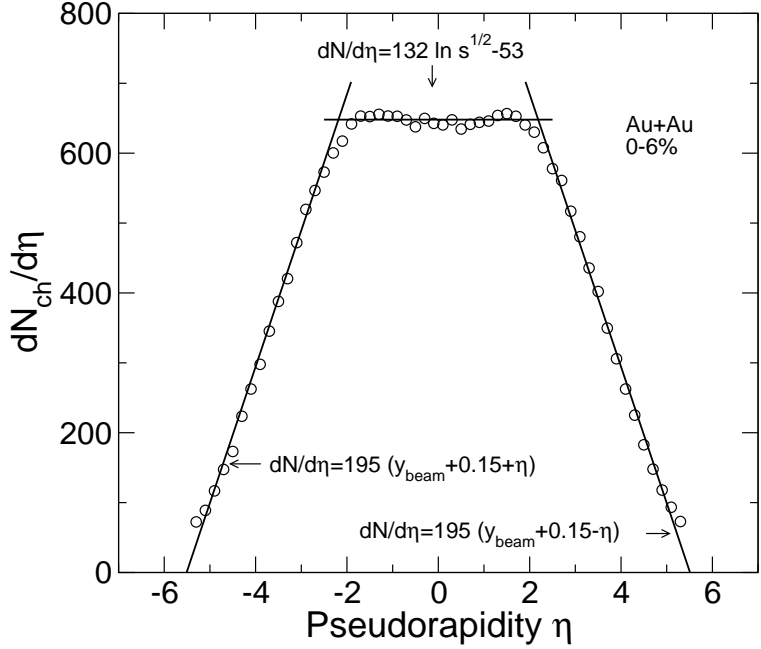}}
\caption{$dN_{ch}/d\eta$ for 0-6\% central 200 GeV $Au+Au$ collisions~\cite{limfrag} (open circles) is seen to be well approximated by a trapezoidal shape (solid lines) for which the parameters are given in the figure.}
\label{trapezoid}
\end{figure}

Earlier, we have noted that shapes of the $dN_{ch}/d\eta$ distributions are to a reasonable approximation represented by a mid-rapidity plateau followed by a nearly linear fall-off in the fragmentation region. It is therefore tempting to approximate this shape by a trapezoidal distribution as illustrated in Fig.~\ref{trapezoid}.  
We have also seen that the mid-rapidity density $dN_{ch}/d\eta|_{|\eta|<1}/\langle N_{\it part} /2 \rangle$ increases logarithmically with $\sqrt{s_{\it NN}}$ (see Fig.~\ref{dNdeta_eta0}), such that the height of the midrapidity plateau increases logarithmically with energy (see Eqn.~\ref{ln(s)}),
which corresponds to $dN_{ch}|_{|\eta|<1}/d\eta=132\ln\sqrt{s_{\it NN}}-53$ for 0-6\% central $Au+Au$ collisions with $N_{\it part}$=344. In addition, the linear part of the limiting fragmentation region is well represented by
\begin{equation}
\frac{2}{\langle N_{\it part}\rangle}\frac{dN_{\it ch}}{d\eta}=1.134(y_{\it beam}+0.15-\eta),
\end{equation}
which for 0-6\% central $Au+Au$ collisions corresponds to $dN_{\it ch}/d\eta=195(y_{\it beam}+0.15-\eta)$ as illustrated in Fig.~\ref{trapezoid}. The area of the trapezoid, {\it i.e.} $N_{ch}$, may then be computed as
\begin{equation}
N_{ch}/\langle N_{\it part}/2\rangle=\frac{2}{\langle N_{\it part}\rangle}\frac{dN_{ch}|_{|\eta|<1}}{d\eta}\left(2y_{\it beam}+0.3-\frac{1}{1.134}\frac{2}{\langle N_{\it part}\rangle}\frac{dN_{ch}|_{|\eta|<1}}{d\eta}\right),
\label{Nch_norm}
\end{equation}
which corresponds to
\begin{equation}
N_{ch}=\frac{dN_{ch}|_{|\eta|<1}}{d\eta}\left(2y_{\it beam}+0.3-\frac{1}{195}\frac{dN_{ch}|_{|\eta|<1}}{d\eta}\right)
\end{equation}
for 0-6\% central $Au+Au$ collisions. 

\begin{figure}
\centerline{\epsfig{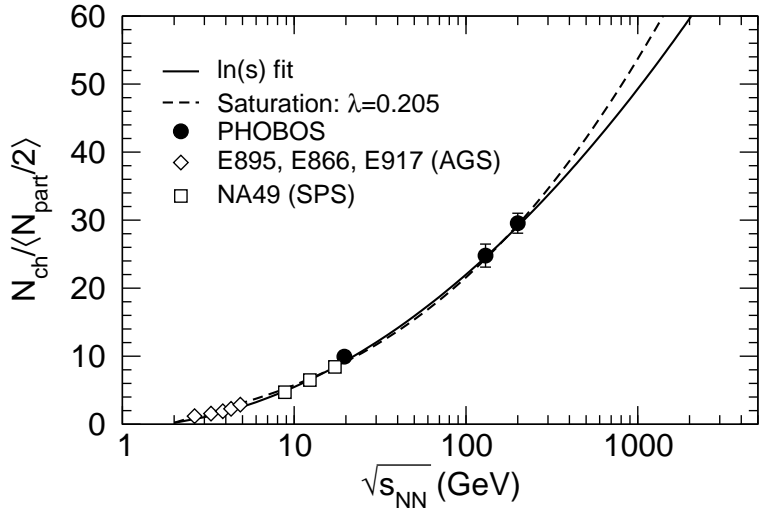}}
\caption{$N_{ch}/\langle N_{part}/2\rangle$ for central  $Au+Au$ and  $Pb+Pb$ collisions  is seen to be well approximated by the expression given in Eqn.~\ref{Nch_norm} (solid curve). The dashed curve corresponds to the gluon saturation model estimate of $dN_{ch}/d\eta$.}
\label{Total_multiplicity}
\end{figure}

As discussed in Sect. 2, the gluon saturation model~\cite{KN} also gives a good representation of the mid-rapidity density $\frac{2}{\langle N_{\it part}\rangle}\frac{dN_{ch}|_{|\eta|<1}}{d\eta}$ at energies above $\sqrt{s_{\it NN}}$=20 GeV. Using this estimate instead we obtain the dashed curve in Fig.~\ref{Total_multiplicity}, which also gives a good account of the total multiplicity. In conclusion we find that the simple trapezoidal representation of the $dN_{ch}/d\eta$ distributions in heavy-ion collisions gives a good representation of the energy evolution of total charged particle production.

\section{Summary and conclusion}

In this work we have attempted to give a summary of the present status of charged particle multiplicity measurements in heavy-ion collisions with special emphasis on the recent measurements obtained at RHIC, but also studying how these results compare with lower energy measurements of $Au+Au$ and $Pb+Pb$ collisions studied at the AGS and SPS, respectively. Parallels and differences between the heavy-ion data and those obtained in nucleon-nucleon, nucleus-nucleon, and lepton-lepton collisions have been discussed. The first result from the new RHIC facility indicated a logarithmic trend in the energy evolution of the mid-rapidity density that was not predicted by most theoretical models. This logarithmic increase with collision energy has been verified in subsequent measurements and has helped to focus the attention on models that approximately reproduce this trend, one such model being based on the gluon saturation concept~\cite{KN,KL,KLN}. The mid-rapidity density in $Au+Au$ collisions shows only a rather weak centrality dependence when  normalized to the number of participant pairs $N_{\it part}/2$; for the most central collision we observe at $\sqrt{s_{\it NN}}$=200 GeV a 40\% enhancement over that for nucleon-nucleon collisions at the same energy. The observed dependence is, however, much weaker than a scaling with the number of nucleon-nucleon collisions would suggest. 

The distributions of charged particles in pseudo-rapidity space exhibit a midrapidity plateau followed by an almost linear fall-off to higher values of $|\eta|$ corresponding to the fragmentation region. A comparison with distributions in rapidity space of identified pions and kaons performed by the BRAHMS collaboration show that the mid-rapidity plateau seen in $\eta$-space should not be interpreted a evidence for a boost invariant region. 

A comparison of $dN_{\it ch}/d\eta$ distributions at three different energies in $Au+Au$ collisions demonstrate that the limiting fragmentation scaling observed earlier in nucleon-nucleon collisions hold rather rigorously also in heavy-ion collisions. Measurements of heavy-ion $dN_{ch}/d\eta$ distributions allow for a small extrapolation to obtain the total number of charged particles emitted from the collision region. It is found that for 200 GeV $Au+Au$ this quantity, scaled by the number of participant pairs, $N_{\it part}/2$, is essentially constant as a function of centrality, but at a level of about 40\% higher than for nucleon-nucleon and deuteron-gold collisions. Finally, we propose a simple expression, which is based on the observed approximate trapezoidal shape of the $dN_{ch}/d\eta$ distributions in $Au+Au$ collisions, for which the height of the mid-rapidity plateau is given by the observed logarithmic dependence on the collision energy and the width is determined by the limited fragmentation scaling. This phenomenological expression provides an excellent account of the total multiplicity in heavy-ion collision over two orders of magnitude in collision energy.

%
%
%
%
\ack
I acknowledge many elucidating discussions with my colleagues in the PHOBOS collaboration. This work was partially supported by U.S. DOE grants 
DE-AC02-98CH10886,
DE-FG02-93ER40802, 
DE-FC02-94ER40818,  
DE-FG02-94ER40865, 
DE-FG02-99ER41099, and
W-31-109-ENG-38, by U.S. 
NSF grants 9603486, 
0072204,            
and 0245011,        
by Polish KBN grant 1-PO3B-062-27(2004-2007), and
by NSC of Taiwan Contract NSC 89-2112-M-008-024.

\smallskip

\end{document}